\documentclass{emulateapj}
\lefthead{Laha et al.}

\usepackage{soul}
\usepackage{times}
\usepackage{url}
\usepackage{graphicx}
\usepackage[usenames]{color}
\DeclareGraphicsExtensions{.pdf,.png,.jpg,.mps,.eps,.ps}
\usepackage{amsmath}
\usepackage{natbib}
\usepackage{tablefootnote}

\def\unit #1{\,{\rm #1}}

\newcommand\kms{\rm \,\unit{km\,s^{-1}}}

\newcommand\cmsqi{\rm \,\unit{cm^{-2}}}

\newcommand\cmcubei{\rm \,\unit{cm^{-3}}}

\newcommand\punit{\rm \,photons\,cm^{-2}\,s^{-1}}

\newcommand\nh{\rm N_{H}}

\newcommand\si{Si\,{\sc ii}\,}

\def\aap{A\&A} \def\apjl{ApJ}  
 \def\apjs{ApJS}

\usepackage{graphicx}

\begin{document}

\title{Ultraviolet emission lines of Si\,{\sc ii} in quasars --- investigating the ``Si\,{\sc ii} disaster"}

\author{Sibasish Laha\altaffilmark{1}, Francis P. Keenan\altaffilmark{2}, Gary J. Ferland\altaffilmark{3}, Catherine A. Ramsbottom\altaffilmark{1},Kanti M. Aggarwal\altaffilmark{2}}\altaffiltext{1}{{Centre for Theoretical Atomic, Molecular and Optical Physics, School of Mathematics and Physics, Queen's University Belfast, Belfast BT7 1NN, Northern Ireland, U.K.}; {\tt email: s.laha@qub.ac.uk}}\altaffiltext{2} {Astrophysics Research Centre, School of Mathematics and Physics, Queen's University Belfast, Belfast BT7 1NN, Northern Ireland, U.K.}\altaffiltext{3} {Department of Physics and Astronomy, The University of Kentucky, Lexington, KY 40506, U.S.A.}

\begin{abstract}

The observed line intensity ratios of the \si 1263 and 1307\,\AA\ multiplets to that of \si 1814\,\AA\ in the 
broad line region of quasars are both an order of magnitude larger than the theoretical values. This was first pointed 
out by Baldwin et al. (1996), who termed it the ``\si disaster", and it has remained unresolved.
We investigate the problem in the light of newly-published atomic data for \si. Specifically, we perform broad line region calculations using several different atomic datasets within the  CLOUDY modeling code
under optically thick quasar cloud conditions. In addition, we test for selective pumping by the source photons or
intrinsic galactic reddening as possible causes for the discrepancy, and also consider blending with other species. 
However, we find that none of the options investigated resolves the \si disaster, { with the potential exception of microturbulent velocity broadening and line blending}. { We find that a larger microturbulent velocity ($\sim 500\kms$) may solve the \si disaster through continuum pumping and other effects}. The CLOUDY models indicate strong blending of the \si 1307\,\AA\ multiplet with emission lines of O\,{\sc i}, although the 
predicted degree of blending is incompatible with the observed 1263/1307 intensity ratios. Clearly, more work is required 
on the quasar modelling of not just the \si lines but also nearby transitions (in particular those of O\,{\sc i}) to fully 
investigate if blending may be responsible for the \si disaster.
\end{abstract}

\keywords{atomic processes, quasars: emission lines }

\vspace{0.5cm}

\section{INTRODUCTION}

Emission and absorption lines of \si provide important diagnostics of the plasma conditions in the low temperature clouds
 of quasars and other active galactic nuclei { \citep[][and references therein]{2002ApJ...570..514D,2009ApJ...706..525M,2011ApJ...739..105S,2012ApJ...751..107B,1997ApJ...489..656L,2001ApJS..134....1V,2007ApJS..173....1L}}. The
 relevant transitions are from the lower-lying energy levels to the ground-state, and lie 
 in the UV wavelength range from $\sim$\,1000 -- 2500\,\AA. \citet{1996ApJ...461..664B} studied \si emission lines arising 
 from the quasar broad line region (BLR) using data from the 4\,m Cerro Tololo Inter-American Observatory (CTIO) telescope,
 and found that the observed ratios of the \si line fluxes at 1263\,\AA\ (3s$^{2}$3p $^{2}$P--3s$^{2}$3d $^{2}$D) and 1307\,\AA\
 (3s$^{2}$3p $^{2}$P--3s3p$^{2}$ $^{2}$S) to that of the 1814\,\AA\ multiplet of \si 
 (3s$^{2}$3p $^{2}$P--3s3p$^{2}$ $^{2}$D) are both more than 
 an order of magnitude larger than the theoretical values. They termed this the ``\si disaster", which forms the main 
 focus of our paper. Since the Baldwin et al. work, there have been several observations of the BLR clouds in narrow-line quasars using high resolution Hubble Space Telescope (HST) data, which show similar discrepancies between theory and observation for the \si emission lines { \citep{1997ApJ...489..656L,2001ApJS..134....1V,2007ApJS..173....1L}.}

  In this paper we address this discrepancy in three ways. Firstly, using 
  recently published \si atomic data by \citet{2014MNRAS.442..388A}, we check if the discrepancy could 
  be due to inaccurate atomic data being adopted by \citet{1996ApJ...461..664B} in their plasma modeling. 
  Secondly, we investigate if continuum pumping by the quasar/active galactic nuclei (AGN) may have a selective 
  effect on the excitation of the 1263 and 1307\,\AA\ multiplet emission lines compared to that at 1814\,\AA. 
  Finally, we assess the impact of blending of the \si transitions with other emission features as a possible source of the 
  discrepancy.

 Our paper is arranged as follows. In \S\ 2 we discuss observations of narrow-line quasars and the discrepancies with theory known as the \si disaster, while in \S\ 3 we describe the new theoretical models. Finally, in \S\ 4
 we provide a discussion of our results.

\section{OBSERVATIONS}

\citet{1996ApJ...461..664B}, in their study of the optical and UV spectrum of the quasar Q\,0207--398, 
detected several emission lines from ions such as O~{\sc i}, N~{\sc v}, O~{\sc vi}, Fe~{\sc ii}, \si and Si~{\sc iii} in the 
(rest-frame) wavelength range 970--2400\,\AA. These emission lines are from the broad line region (BLR) of the quasar, where the plasma is typically photoionized by the incident AGN radiation. Using the measured line intensities and their ratios, Baldwin et al.
constrained the ionising photon flux and the density of the BLR cloud through a comparison with 
theoretical simulations. From the ionization state of the cloud defined by $U=\Phi_{\rm H}/(n_{\rm H} c)$, where $\Phi_{\rm H}$ 
is the incident photon flux, $n_{\rm H}$ is the hydrogen density and $c$ the speed of light,
one can determine the location $R$ of the cloud by knowing $\Phi_{\rm H}$ and the total photon flux $Q_{\rm H}$ of the source, 
where $\Phi_{\rm H}=Q_{\rm H}/4\pi R^2$. Baldwin et al. 
extensively spanned the parameter space of $\Phi_{\rm H}$ and $n_{\rm H}$ and found that the regions which best describe the various line ratios from several ions are unable to reproduce the observed \si emission line intensities, even though their similar Doppler width ($\sim 1000\kms$) point to a common region of origin. In particular, the ratios of the observed \si 1263 and 1307\,\AA\ 
multiplet line intensities to that of \si 1814\,\AA\ were both more than an order of magnitude larger than the theoretical values.
This problem was referred to as the ``\si disaster" by Baldwin et al., who discussed several possibilities for this anomaly, such as the effects of dielectronic recombination, charge transfer, collisional excitation and 
selective excitation, but could not resolve the issue. In Table 1 we list the \si line intensity ratios 
measured by Baldwin et al. for QSO 0207--398, and below discuss a few more instances of quasars which exhibited such \si emission line discrepancies.

The narrow-line quasar I~Zw~1 had been studied several times over the last 20 years in the optical and UV wavelength bands 
\citep[][]{1997ApJ...489..656L,2001ApJS..134....1V,2004A&A...417..515V}. It has been paid such attention 
because its narrow line profiles show minimal blending, thus allowing emission lines to be individually identified and measured. \citet{1997ApJ...489..656L} observed this source using the Faint Object Spectrograph (FOS) on board HST, and detected
the \si emission line multiplets at 1263, 1307 and 1814\,\AA.  However, the authors pointed out 
a possible blend of O~{\sc i} with the \si multiplet at 1307\,\AA. Table 1 lists the \si emission line ratios measured by \citet{1997ApJ...489..656L}.

\citet{2001ApJS..134....1V} similarly studied I~Zw~1 with FOS/HST, with the aim of providing an empirical UV template for Fe emission in quasars. Their \si line intensity ratios are also reported in Table 1. 

Optical and UV spectra of the quasar PHL\,1811 were obtained by \citet{2007ApJS..173....1L} using the Space Telescope Imaging Spectrograph (STIS) on board HST, plus the 2.1\,m telescope at the Kitt Peak National Observatory. This is a narrow line quasar 
whose UV spectrum is dominated by Fe~{\sc ii} and Fe~{\sc iii} lines, and unusual low ionization species such as 
Na~{\sc i} and Ca~{\sc ii}. The higher ionization stage emission lines are very weak, which Leighly et al. 
attribute to an unusually soft spectral energy distribution. They detected the \si emission lines 
in the UV spectrum, and their intensity
ratios are summarised in Table 1.

In all the above cases we find that the observed fluxes for the \si multiplets at 1263 and 1307\,\AA\ are $5-10$ times larger than 
that of \si 1814\,\AA. By contrast, 
the simulations predict larger fluxes for the multiplet at 1814\,\AA,  as discussed in \S\ 3.

\section{CLOUDY MODELS}

We have used the photoionization code CLOUDY \citep{1998PASP..110..761F,2013RMxAA..49..137F} for our modeling work, which
was also adopted by \citet{1996ApJ...461..664B}. The CLOUDY models generated by Baldwin et al. employed \si 
transition probabilities (TP) from the compilation of \citet{1988ApJS...68..449M}, and the results of 
\citet{1991MNRAS.248..827D} for electron impact excitation effective collision strengths (ECS). Over the last 20 years, the 
available data for these atomic processes have been improved, and the most recent release of CLOUDY (Ferland et al. 2013) 
employs \si TP and ECS  
values from \citet{1998ADNDT..68..183N} and \citet{2008ApJS..179..534T}, respectively. However, very recently 
\cite{2014MNRAS.442..388A} have produced new calculations of TP and ECS for \si, considering all 1540 transitions 
among the lowest 56 energy levels. These are estimated to be accurate to $\pm$\,20\%\ for most transitions, and in some instances are very different from previous work. For example, for the 3s$^{2}$3p $^{2}$P$_{1/2}$--3s3p$^{2}$ $^{2}$D$_{3/2}$ (1808.01\,\AA) transition
at an electron temperature of T$_{e}$ = 10,000\,K, the Tayal (2008) value of ECS = 2.74, about 40\%\ larger than that of Aggarwal \&
Keenan (ECS = 1.91). Similarly, the Nahar (1998) TP for 3s$^{2}$3p $^{2}$P$_{3/2}$--3s$^{2}$3d $^{2}$D$_{5/2}$ (1264.73\,\AA)
is 3.04$\times$10$^{9}$\,s$^{-1}$, over 30\%\ larger than the Aggarwal \& Keenan calculation of 2.31$\times$10$^{9}$\,s$^{-1}$. For some transitions, the differences in TP are even larger, such as 
3s$^{2}$3p $^{2}$P$_{1/2}$--3s3p$^{2}$ $^{2}$D$_{3/2}$, where the Nahar calculated value is more than a factor of 10 
greater than that of Aggarwal \& Keenan (2.54$\times$10$^{6}$\,s$^{-1}$ compared to 1.0$\times$10$^{5}$\,s$^{-1}$). { See Table 1 of \citet{2016MNRAS.455.3405L} for a comparison of the TP and ECS values between the various atomic datasets of \si.}

In view of the above, we investigate if the ``\si disaster'' anomaly may be due to the adoption of inaccurate atomic data. 
Specifically, we have created three CLOUDY models with differing atomic datasets. The first (termed CLOUDY1) employs the same \si TP and ECS as \citet{1996ApJ...461..664B}, i.e. those from \citet{1988ApJS...68..449M} and \citet{1991MNRAS.248..827D}, while the second (CLOUDY2)
is the \citet{2013RMxAA..49..137F} CLOUDY model with the atomic data of \citet{1998ADNDT..68..183N} and \citet{2008ApJS..179..534T}.  In the third (CLOUDY3) we adopt the TP and ECS of Aggarwal \& Keenan (2014). All three models consist of the energetically-lowest 148 fine-structure levels
of \si, with energies from the NIST database.\footnote{http://www.nist.gov/pml/data/asd.cfm} However, the calculations of 
Aggarwal \& Keenan only consider the lowest 56 fine-structure levels. Hence CLOUDY3 is a merger of the datasets of
Aggarwal \& Keenan and CLOUDY2, where we use the results of the former for the first 56 levels and data from the latter for the remainder. The TP values from Aggarwal \& Keenan were wavelength-corrected to the NIST observed wavelengths.

For each CLOUDY model we have calculated the \si emission line strengths in a BLR cloud. \citet{1996ApJ...461..664B} generated
grids of CLOUDY models covering a large range of hydrogen density ($10^7 \leq n_{\rm H} \leq 10^{14} \cmcubei$) 
and ionizing photon flux ($10^{17} \leq \Phi_{\rm H} \leq 10^{24}\punit$). They used contour plots of these parameters to determine
values which could produce the observed spectrum, and for component A in Q0207--398 found that the \si lines are emitted
in a BLR cloud with $n_{\rm H}=10^{12.7}\cmcubei$ and $\Phi_{\rm H}=10^{20.7}\punit$. These parameters are hence used as { representative values} 
to model the BLR clouds in our CLOUDY simulations, and in Table 1 list the resultant theoretical \si line intensity 
ratios. We note that the ``stopping" criterion for the CLOUDY calculations is when the total hydrogen column density ($\nh$) of the cloud reaches $10^{23}\cmsqi$, which yields the optically thick case. { The equivalent width of the \si $1814 \rm \AA$ { multiplet} ($W_{\lambda}\sim 1.81 \rm \AA$) observed { in the quasar Q0207-398} by \citet{1996ApJ...461..664B} compares well with that calculated using CLOUDY ($W_{\lambda}\sim 2.11 \rm \AA$), using the BLR parameters mentioned above, and a unit cloud covering fraction.} 
{ We note that we have assumed a Solar metallicity in the above calculations, but consider non-Solar values in Section 4.}

 
\section{Results and discussion}

As noted in \S\ 3,  the recent TP for the \si 1814\,\AA\ (3s$^{2}$3p $^{2}$P--3s3p$^{2}$ $^{2}$D) multiplet lines
calculated by Aggarwal \& Keenan (2014) are more than a factor of 10 smaller than the earlier values of Nahar (1998). This would hence appear to potentially provide an explanation for the \si disaster, as reducing the TP for the \si 1814\,\AA\ multiplet
might be expected to similarly reduce the theoretical line intensity, hence increasing the predicted values of the 
1263/1814 and 1307/1814 ratios, hopefully to match the observations and hence solve the \si disaster problem. However, from Table 1
we see that the observed values of the 1263/1814 and 1307/1814 ratios range from 2.1--6.6 and 2.8--5.7, respectively, while the CLOUDY3 calculations (which use the Aggarwal \& Keenan TP data) are 0.63 and 0.40, respectively. These theoretical values are significantly larger than those from CLOUDY1 and CLOUDY2, but still not by a sufficient amount to resolve the \si { emitted spectrum} discrepancy. { Hence we conclude that the latest atomic data do not solve the \si disaster}. 

\begin{figure}
  \centering

\includegraphics[width=9cm,angle=0]{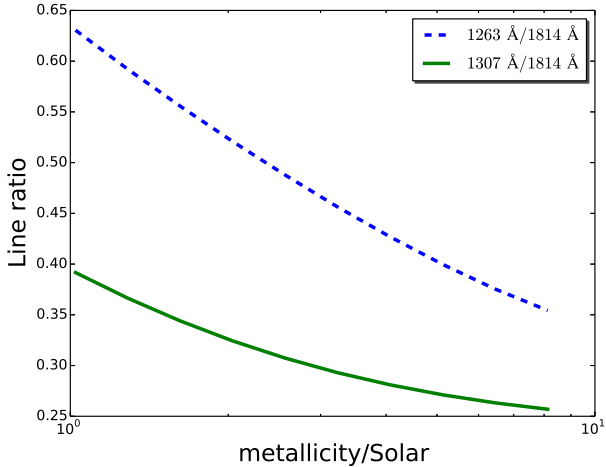}

\caption{\si line ratios plotted as a function of cloud metallicity (in units of the Solar value).} \label{fig:metals}
\end{figure}

{ \citet{1996ApJ...461..683F} found that the BLR clouds may have super-solar metallicities ($\sim$\,5\,Z$_{\odot}$), which can change the ionic column densities and hence the optical depths of different emission lines, in turn affecting the line ratios. Figure \ref{fig:metals} shows the \si emission line ratios calculated as a function of cloud metallicity, which is varied from 1--10 times Solar. We find that the resultant theoretical line ratios decrease with increasing metallicity of the cloud, and are in worse agreement with observation. Hence, enhancing the metallicity of the cloud does not solve the \si disaster.}


However, another possible explanation is continuum pumping. 
In AGN, the continuum photoionizes the BLR gas clouds and can selectively pump specific levels and hence lines. 
The optimal way to test if the \si 1263 and 1307~\AA{} lines are selectively pumped by the continuum would be to switch off and on the continuum and compare the intensities. However, the emission line strength are dependent on the ionization and thermal equilibrium of the BLR cloud, and upon changing the spectral energy distribution or switching it off, the equilibrium will be disturbed and the line ratios will change not only because of photoionization, but also other effects. Therefore, we have adopted 
an alternative method to test this effect, by first reducing the number of available levels of \si in the CLOUDY3 model to 
11 (the minimum number required to produce the \si emission lines), and then allowing all 148 levels to be in use. 
By comparing the predicted line intensities in the two instances we can estimate the effect of indirect photoexcitation, { whereby the \si electrons are pumped to higher levels by the continuum and then cascade to strengthen the lines of  interest.} However, we note that in all cases the \si line fluxes changed by $\leq$\,5\%, and thus continuum pumping by indirect photoionization cannot be a possible solution to the \si problem. 

The microturbulent velocity of a cloud is also a potential source of continuum pumping. Turbulence broadens the local line width, which can then absorb a larger fraction of the continuum, leading to increased line intensity. Also, the presence of turbulence in a cloud reduces the optical depth and hence increases the line intensities, as the emitted photons can escape more easily. 
The effect of microturbulent velocity on BLR clouds have been studied extensively by \citet{2000ApJ...542..644B} using CLOUDY. These authors found that the \si line multiplets at $1263 \rm \AA$ and $1307 \rm \AA$ are selectively pumped by the continuum to a far
greater extent than the $1814 \rm \AA$ multiplet, for turbulent velocities ranging from $100-10^4 \kms$. By default, CLOUDY 
adopts a microturbulent velocity of $0 \kms$, and hence we have undertaken calculations 
with CLOUDY3 data for a turbulent velocity of $500 \kms$, and derived $1263/1814$ and $1307/1814$ ratios of $4.80$ and $2.62$, respectively. These values are much larger than the results for a turbulent velocity of $0\kms$, and closer to the observed ratios (see Table 1). { Therefore the microturbulent velocity broadening of the BLR clouds could be a possible solution to the \si disaster.} However, there is an important caveat to this exercise. The introduction of turbulent velocity into a cloud changes 
its entire properties, including temperature and ionisation structure. Hence { it is hard to isolate} the effect of continuum pumping 
on the emission lines, as several other factors also affect the line emissivity. 

A potential source of the \si discrepancy could be intrinsic reddening by Galactic-like dust which produces a pronounced broad absorption feature in the range 1800--2500\,\AA, which reduces the intensity of the \si 1814\,\AA\ feature and
hence leads to enhancements in the 1263/1814 and 1307/1814 ratios over their true values and the \si disaster. 
\citet{1997ApJ...489..656L} discuss this effect in detail for I\,Zw\,1, but note there was little evidence for the presence of such 
intrinsic reddening. This therefore appears to be an unlikely cause for the \si problem.

In a typical BLR plasma, the clouds are in a Keplerian orbit about the supermassive black hole at velocities of 
$\sim$\,1000\,km\,s$^{-1}$, which would lead to line broadening of approximately 4 and 6\,\AA\ at 1300 and 
1800\,\AA, respectively. Hence the \si disaster may simply be due to 
blends, as found for other such long-standing problems. For example, \citet{1990ApJ...353..323D} found discrepancies between theory and 
observation for emission lines of Fe~{\sc xv} in solar flares, which they attributed to either errors in the adopted atomic data or the 
effects of atomic processes which were not considered in their flare models. However, subsequently \citet{2006A&A...449.1203K}
showed that line blending was responsible. To investigate this for \si, we have used CLOUDY to 
calculate the intensities of possible blending lines in the wavelength ranges 1258.4--1267.0\,\AA\ (i.e.
spanning the components of the 1263\,\AA\ multiplet, plus $\pm$\,2\,\AA), 1302.4--1311.3\,\AA\ (the same for the 1307\,\AA\
multiplet), and 1805.0--1820.5\,\AA\ (spanning the components of the 1814\,\AA\ multiplet, plus $\pm$\,3\,\AA\ in this instance).
We list the calculated intensities of the blending lines in Table 2 relative to that of the relevant \si multiplet. Only lines which 
are predicted to be  $\sim$\,5\%\ or greater
of the intensity of the \si feature are included in the table, and we note that no blending lines were
found for the 1814\,\AA\ multiplet in any of the CLOUDY models. Also shown in Table 2 are the revised theoretical \si line ratios taking into account the effect of the blends. An inspection of the table reveals that the revised values of 1307/1814 are now closer to
the {observational} results, with the CLOUDY3 theoretical ratio (3.6) being in reasonable agreement with the observations (which range from 2.8--5.7).
However, the revised estimates for 1263/1814 remain significantly lower  than the measured values. In addition, { the predicted
ratios for the line blend flux ratios of} 1263/1307 are 0.13 (CLOUDY1), 0.21 (CLOUDY2) and 0.20 (CLOUDY3), much smaller than the observed values of
0.51--1.9. This discrepancy arises due to the prediction of very strong O\,{\sc i} emission in the CLOUDY models, with in fact this species 
dominating the 1307\,\AA\ feature. 
The O\,{\sc i} 
lines are excited by the H\,{\sc i} Ly$\beta$ Bowen fluorescence process, which we treat as in \citet{1985ApJ...291..464E}. This process is complex and depends on the detailed velocity and density structure of the plasma.
Given the complexity in dealing with O\,{\sc i}, it is possible (and indeed
perhaps likely) that the predicted O\,{\sc i} intensities are not reliable, so that our estimates of line blending are in turn 
not highly accurate. There is also no {\em a priori} reason to believe that our calculations of blends for the 1263 and 1814\,\AA\ 
multiplets are reliable. We therefore conclude that blending cannot be ruled out as a source of the \si discrepancy, but clearly 
more work is required on the calculation of the intensities, and hence impact, of blending species.


In summary, we have ruled out several possible explanations for the \si disaster observed in quasar spectra, including 
errors in atomic data, continuum pumping  and the presence of intrinsic reddening in the source. { We find that an enhanced microturbulent velocity in the BLR plasma can solve the \si disaster. However, the caveat is that changing the turbulent velocity also changes the ionic structure and several other properties of the cloud, and hence the effect of continuum pumping on the line ratios may not be isolated. Another possible explanation for \si disaster, line blending, cannot we believe be completely ruled out at this stage,} and more detailed calculations of the intensities of possible blending species are required.

\section*{ACKNOWLEDGEMENTS}

 The project has made use of public databases hosted by SIMBAD, maintained by CDS, Strasbourg, France. SL, CAR and FPK are grateful to STFC for financial support via grant ST/L000709/1. GJF acknowledges financial support from the Leverhulme Trust via Visiting Professorship grant VP1-2012-025, and also support by the NSF (1108928, 1109061 and 1412155), NASA (10-ATP10-0053, 10-ADAP10-0073, NNX12AH73G and ATP13-0153) and STSciI (HST-AR-13245, GO-12560, HST-GO-12309, GO-13310.002-A, HST-AR-13914 and 
 HST-AR-14286.001). { We thank the referee, Prof. Kirk Korista, for insightful comments which helped us to improve the  manuscript.}


\begin{table*}

\centering
\begin{minipage}{140mm}
\caption{Observed and theoretical \si emission line intensity ratios} \label{}
  \begin{tabular}{lccccccccccccccc} \hline\hline 
	
  Ratio		& Q\,0207--398$^1$ & I\,Zw\,1$^{2}$ & I\,Zw\,1$^{3}$  & PHL\,1811$^{4}$ & CLOUDY1$^{5}$	& CLOUDY2 	& CLOUDY3$^{6}$ 	 & CLOUDY3$^{7}$	
  \\ 
  \hline
\\
1263/1814	& 6.6 & 5.2  & 2.1 & $\geq$\,8.5		& 0.15 	& 0.20 &  0.63  &4.80	
\\  
1307/1814 	& 5.7 	& 2.8 & 4.1 & $\geq$\,2.4 		& 0.17  & 0.12 & 0.40 	&2.62
\\ 
 \hline \hline
\end{tabular} 

$^1$From \citet{1996ApJ...461..664B}
\\
$^{2}$From \citet{2001ApJS..134....1V}
\\
$^{3}$From \citet{1997ApJ...489..656L}
\\
$^{4}$From \citet{2007ApJS..173....1L}
\\
$^{5}$See \S\ 3 for details of the different CLOUDY models.
 
$^{6}$Calculations for microturbulent velocity $= 0 \kms$.

$^{7}$Calculations for microturbulent velocity $= 500 \kms$.

\end{minipage}

\end{table*}


\begin{table}
\begin{minipage}{100mm}
\caption{Theoretical intensity ratios for possible blending lines} \label{}
  \begin{tabular}{lccc} \hline\hline 
	
Line ratio					& CLOUDY1 & CLOUDY2 & CLOUDY3  	
\\
\hline
S\,{\sc ii} 1259.52/\si 1263    & 0.120 &  0.107 & 0.104 
\\
Si\,{\sc i}/\si 1263 & 0.049 & 0.044 & 0.043 
\\
P\,{\sc ii} 1301.87/1307 & 0.065 & 0.073 & 0.073 
\\  
O\,{\sc i} 1302.17/\si 1307 & 2.3 & 2.6 & 2.6 
\\  
Si\,{\sc iii} 1303.32/\si 1307		&  0.50 & 0.57 & 0.57 
\\  
P\,{\sc ii} 1304.49 /\si 1307	&  0.066 & 0.075 & 0.075 		
\\  
P\,{\sc ii} 1304.68/\si 1307	& 0.061 & 0.069 & 0.069 	
\\  
O\,{\sc i} 1304.86/\si 1307	& 2.1 & 2.4 & 2.3 		
\\  
P\,{\sc ii} 1305.50/\si 1307 & 0.065 & 0.073 & 0.073  		
\\  
O\,{\sc i} 1306.03/\si 1307	&  1.7 & 2.0 & 1.9 		
\\  
P\,{\sc ii} 1309.87/\si 1307	& 0.072 & 0.082 & 0.082		
\\
1263/1814$^{1}$  & 0.18 & 0.23 & 0.72
\\
1307/1814$^{1}$  & 1.4 & 1.1 & 3.6
\\  
\hline \hline
\end{tabular} 

$^1$Revised theoretical values of line ratios with the addition of the predicted blending lines.

\end{minipage}

\end{table}

\bibliographystyle{apj}

\bibliography{mybib}

\end{document}